\documentclass[twocolumn,prl,showpacs]{revtex4-1}
\usepackage{epsf,graphicx,amssymb,color,upgreek}

\begin{document}
\title{Drop impact of shear thickening liquids}
\author{Fran\c cois Boyer}
\author{Jacco H. Snoeijer}
\author{J. Frits Dijksman}
\author{Detlef Lohse}
\email[E-mail: ]{d.lohse@utwente.nl}
\affiliation{Physics of Fluids Group, MESA+ Institute and Faculty of Science and Technology, University of Twente, P.O. Box 217, 7500 AE Enschede, Netherlands}
\date{\today}

\begin{abstract}  
The impact of drops of concentrated non-Brownian suspensions (cornstarch and polystyrene spheres) onto a solid surface is investigated experimentally. The spreading dynamics and maximal deformation of the droplet of such shear thickening liquids are found to be markedly different from the impact of Newtonian drops. A particularly striking observation is that the maximal deformation is independent of the drop velocity and that the deformation suddenly stops during the impact phase. Both observations are due to the shear-thickening rheology of the suspensions, as is theoretically explained from a balance between the kinetic energy and the viscously-dissipated energy, from which we establish a scaling relation between drop maximal deformation and rheological parameters of concentrated suspensions.

\end{abstract}

\maketitle

The impact of liquid drops on solid surfaces is a classical experiment in interfacial hydrodynamics \cite{Worthington} and has received an increasing attention from a broad community (from applied mathematics to soft matter physics to chemical engineering) \cite{Yarin2006}. Impact is relevant for a large number of industrial processes (e.g. inkjet-printing \cite{Wijshoff2010}, spray coating, pesticide delivery \cite{Bergeron2000}) and touches upon the fundamental challenges of spreading, splashing and bouncing that arise in this paradigmatic experiment. Recent studies have been dedicated to provide a better understanding of the spreading dynamics, in particular on superhydrophobic \cite{Clanet2004},  superheated \cite{Biance2011} and/or microstructured surfaces \cite{Tran2012a, Tran2012b}, and used a variety of experimental techniques (Fourier-transform profilometry \cite{Lagubeau2012}, particle image velocimetry \cite{Lastakowski2013}, high-speed color interferometry \cite{Veen2012}) to get new insights into the complex hydrodynamics of an impacting droplet. 
Splashing remains the least understood regime as it has been revealed to depend on a large number of parameters  \cite{Range98,Tsai, Bird2009} and involves an intricate coupling between air and liquid flows \cite{Xu2005,Bouwhuis2012}.

Despite the complex mechanical response of a large class of fluids used in relevant applications (paint, blood for forensics studies, and colloids for the increasingly popular 3D-inkjet-printing \cite{Lewis2006}, most drop impact studies have focused on Newtonian liquids. However, as we will show in this paper, non-Newtonian properties affect the drop impact dynamics considerably.Vice versa, such impact studies may be applied to accurately extract the complex rheology of such materials. It was already shown that visco-elastic properties have a strong influence on the retraction dynamics after impact \cite{Bartolo2007}. More recently, yield-stress and shear-thinning fluids have been extensively studied and successful scaling laws have been introduced to relate the rheological properties to the drop maximal deformation \cite{Luu2009,Guemas2012}. 

\begin{figure}[htp]
                    \center{\includegraphics[width=8.6cm]{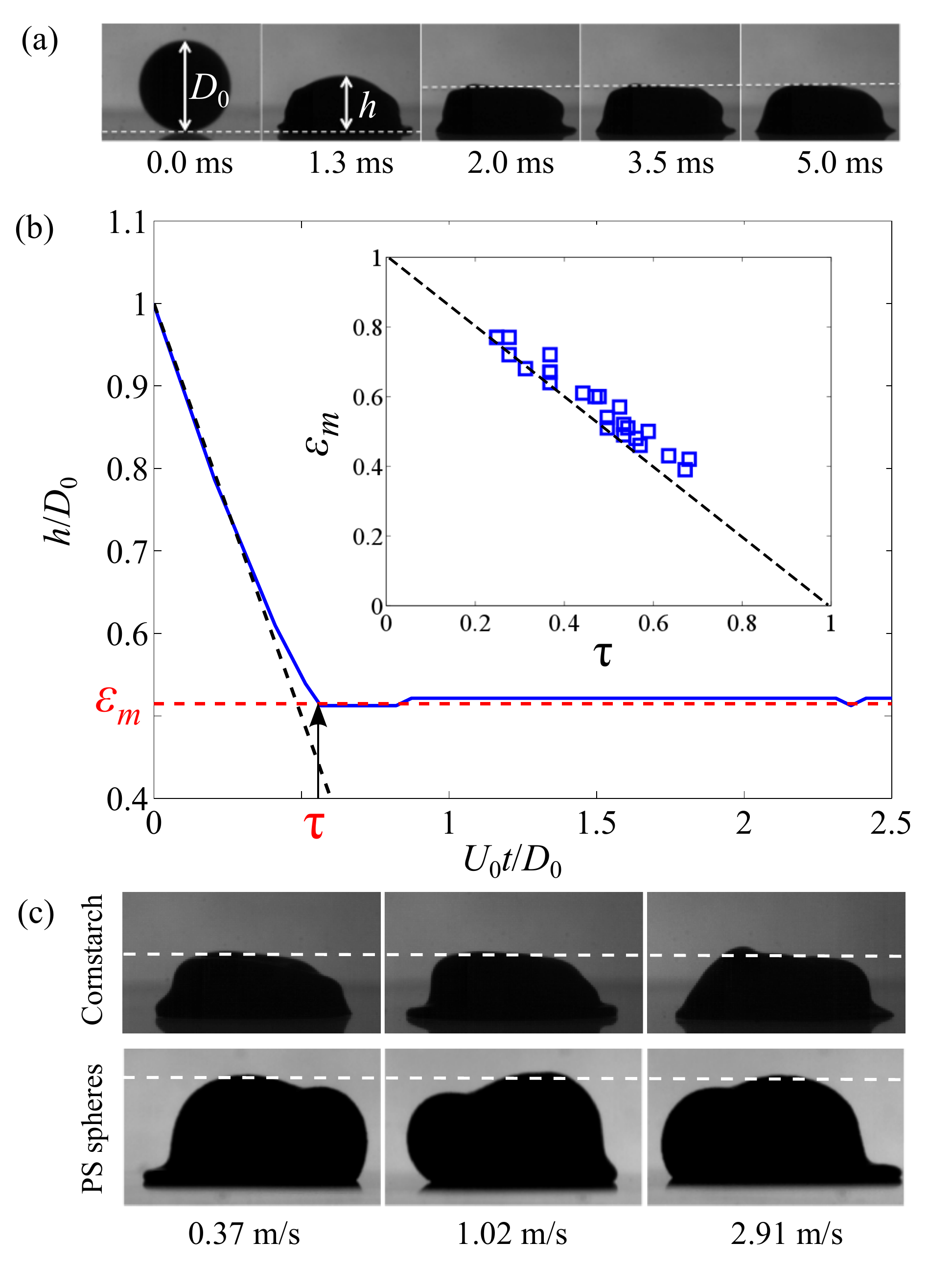}}
                    \caption{(color online) (a) Image sequence of a drop of cornstarch suspension impacting on a glass plate ($D_0= 2.80$~mm, $U_0=1.04$~m/s, $\phi_0=35\%$). (b) Dimensionless drop height $h/D_0$  as a function of dimensionless time $U_0 t/D_0$; kinematic (dashed black line) and \emph{frozen} states (red dashed line), defining the plateau value $\varepsilon_m$, which is shown in the inset as function  of the dimensionless crossover time $\uptau$,  also defined in (b). (c) Snapshots of  \emph{frozen} drops of cornstarch ($D_0=2.80$ mm, $\phi_0=35\%$), and PS spheres ($D_0=2.23$~mm, $\phi_0=55\%$) suspensions, for different impact velocities. At later times, the drops will slowly spread and recover an axisymmetric shape.  See Supplemental Material at [URL] for the corresponding movies (cornstarch, $\phi_0=35\%$: $U_0 = 0.37$ m/s, $U_0 = 1.02$~m/s, and $U_0 = 2.91$ m/s,).}  \label{fig1}
                    \end{figure}

In this Letter, we investigate drop impact for concentrated suspensions, which are known to exhibit shear-thickening effects. We reveal that during the impact phase, the liquid ``freezes'' into a strongly deformed state (Fig.~\ref{fig1}). Remarkably, this maximum deformation turns out to be completely independent of the impact velocity. We systematically study the parametric dependence of maximal deformation with the drop  size and particle concentration and derive scaling arguments based on the balance between kinetic energy and viscously-dissipated energy. Rheological measurements are used to characterize the shear-thickening regime and provide scaling laws that are in excellent agreement with all observations, such as the independence of the spreading on the impact velocity. 

The experimental setup has been previously described by \cite{Guemas2012}. A drop of suspension is slowly generated out of a capillary needle driven by a syringe pump until it detaches  and falls under its own weight. The released drop has a diameter close to the capillary length and quickly relaxes to a spherical shape. Changing the needle's outer diameter allows to control the falling drop diameter $D_0$ from 1.8 to  3.5~mm. The drop impact velocity $U_0$ is varied by releasing the drop from different heights, in the range of 0.3 - 3.0~m/s. The impacted surface is a clean and smooth microscope glass slide. The impact dynamics is visualized from the side with a high-speed camera (Photron Fastcam APX-RS) 
 at 20,000 fps with a macro-lens. 
 The intense and homogeneous backlighting needed is obtained with a 42W Xenon light source and a light diffuser.

Two types of suspensions are used to perform the experiments: (i) commercially available cornstarch (Duryea, mean diameter $\bar{d}= 5- 20$ ~$\mu$m, density $\rho_p = 1.683$~g/cm$^{3}$), and (ii) polystyrene spherical (PS) particles (Microbeads, diameter $d=20$ $\mu$m, density $\rho = 1.051$~g/cm$^{3}$). Both are dispersed in density-matched aqueous solutions of cesium chloride (dynamic viscosity, $\eta_f=1$~mPa.s ) \cite{Brown2009,Fall2010}. The suspensions were prepared by slowly pouring solid particles into the carrier fluid and vigorously stirring the mixture with a spatula. The samples were finally sonicated for a few minutes before use in the experiments for even better mixing. Even at longer timescales (hours), no settling or creaming of the particle phase was observed. Control of the volume fraction $\phi_0$ (defined as the ratio of the solid particle volume to the total suspension volume) was achieved by weighing both particles and fluid. Cornstarch suspensions were prepared at three different volume fractions: 33\%, 35\% and 38\%, and a suspension of PS spheres was prepared at 55\%.  For both cornstarch and PS particles, the large size of the suspended particles ensures that Brownian motion and colloidal interactions can be safely neglected, and such systems are thus called \emph{granular suspensions}. The use of particles of two chemically-different materials and sphericity allows to test the universality of the drop impact behavior. 

Figure \ref{fig1}  shows snapshots of a typical drop impact experiment (35\% cornstarch suspension, $D_0= 2.80$~mm, $U_0=1.04$~m.s$^{-1}$) and illustrates the peculiar behavior of granular suspensions. In sharp contrast to simple fluids \cite{Lagubeau2012, Lastakowski2013,Eggers2010}, the spreading phase of the impacting drop abruptly stops shortly after impact. The drop is then "frozen" in a deformed state and no macroscopic deformation of the drop interface is observed at intermediate timescales. At longer times ($t > 40$ ms in the present experiment), the drop relaxes to a spherical cap shape: this slow relaxation dynamics (not shown here) depends on the surface properties of the suspension and the substrate but does not depend on the impact parameters ($U_0$, $D_0$) and will not be documented in this Letter. In the following, we will focus on the short-timescales impact dynamics, i.e. at times comparable to the characteristic impact timescale $t_0=D_0/U_0$. The image sequence shown in Fig.  \ref{fig1} also emphasizes a characteristic feature of the drop impact dynamics of suspensions: whereas the so-called \emph{spreading ratio} is seen to reach a maximum value $D_{max}/D_0$ ranging typically from 2 to 10 for Newtonian fluids, the diameter of the base area here remains in the order of $D_0$, i.e. $D_{max}/D_0 \sim 1$. Furthermore, spreading of suspension drops is seen to exhibit a remarkable breaking of the azimuthal symmetry. As a consequence, $D_{max}$ depends on variations of the wetted area at the instant of contact -- the drop height $h$ is chosen as a more suitable quantity to characterize drop deformation upon impact.

From image analysis, the height $h$ of the drop (at apex) is determined for each frame of the high-speed movie and a characteristic time-evolution is shown in Fig.  \ref{fig1}, in dimensionless units $h/D_0$ vs. $U_0 t/D_0$. It can be described as  a two-stage process: (i) a first \emph{kinematic} stage, where $h$ is seen to decrease linearly with time at a constant speed $U_0$ (black dashed line, $h/D_0=1-U_0 t/D_0$), as if the motion of the drop apex was not altered by the impact event; (ii) a \emph{frozen} state, where $h$ stays constant at a plateau value $\varepsilon_m D_0$ for times longer than $D_0/U_0$. The crossover time $\tau$ is defined as the first moment where $h$ reaches the plateau value. An ideally sharp transition between the two stages would then lead $\varepsilon_m = 1- \tau$, which is reasonably verified (within 10\%) in the inset of Fig.  \ref{fig1}. 

    \begin{figure}[htp]
                    \center{\includegraphics[width=8.6cm]{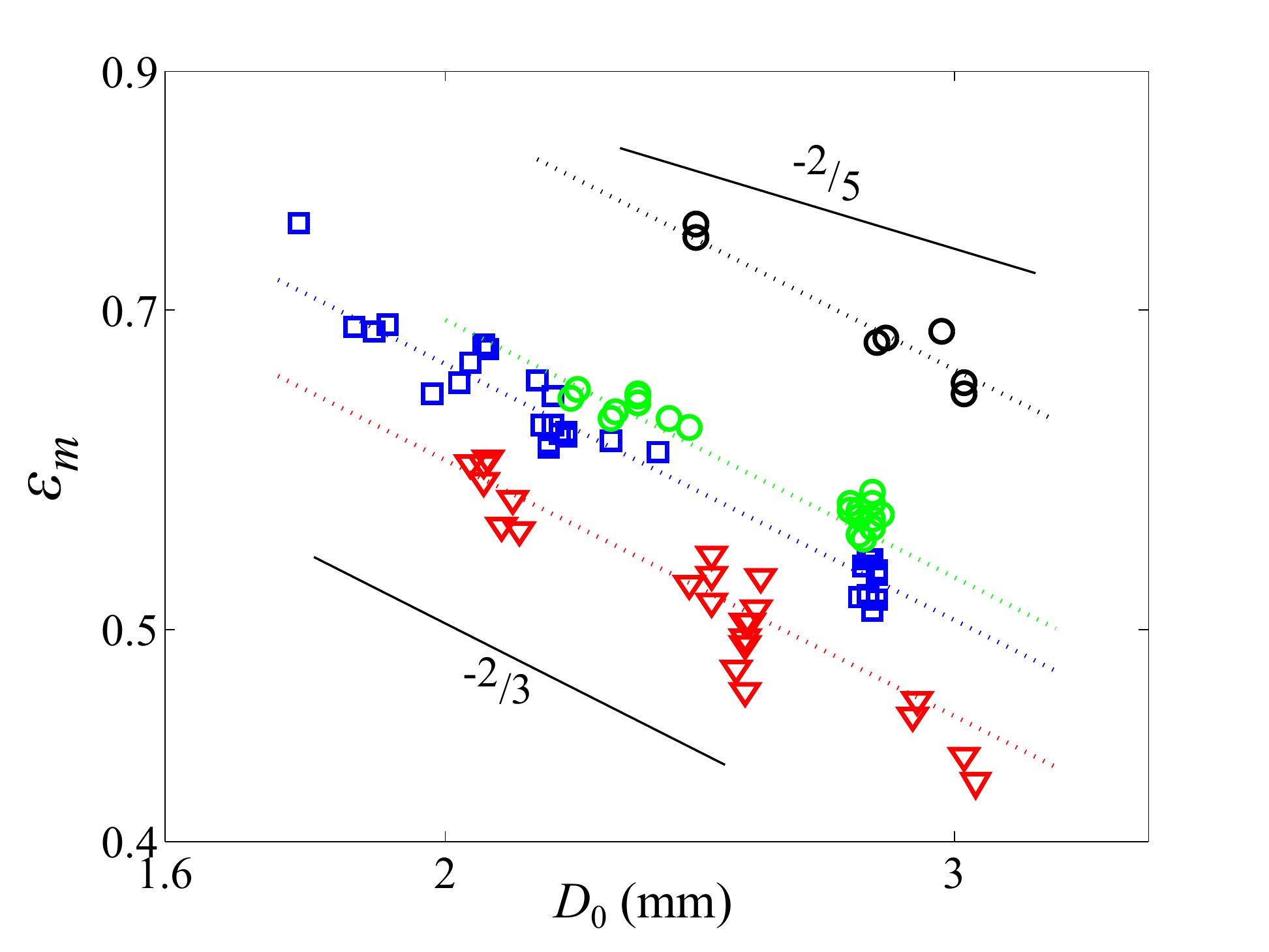}}
                    \caption{(color online) Dimensionless height $\varepsilon_m$ as a function of drop diameter $D_0$ (logarithmic scale), for different cornstarch volume fractions: 33\% (\textcolor{red}{$\bigtriangledown$}), 35\% (\textcolor{blue}{$\square$}), and 38\% ({\scriptsize$\bigcirc$})  ; and for 55\% PS spheres suspension (\textcolor{green}{\scriptsize$\bigcirc$}). 
                    }  \label{fig2}
                    \end{figure}

\begin{figure}[htp]
                    \center{\includegraphics[width=8.6cm]{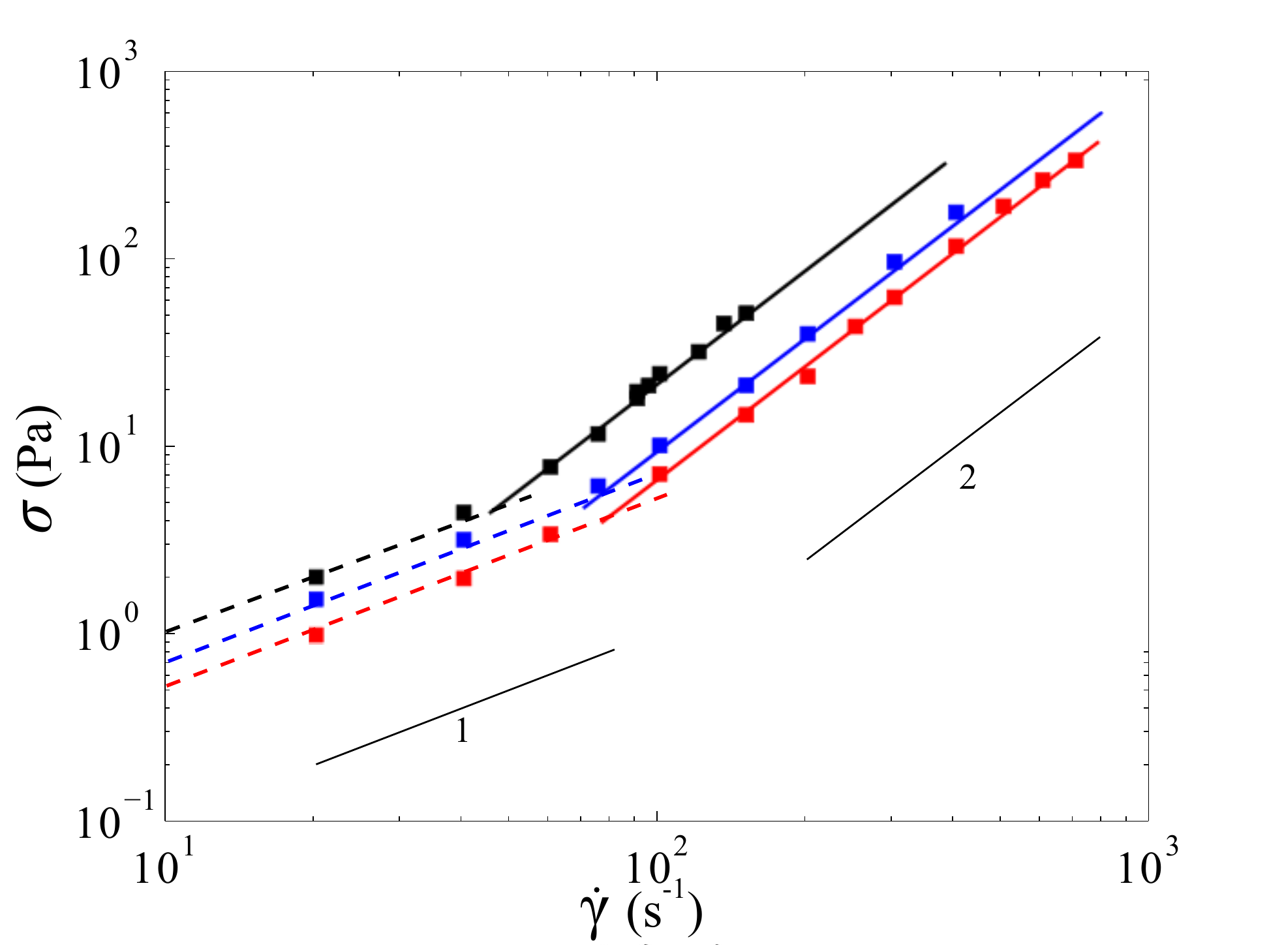}}
                    \caption{(color online) Flow curves of cornstarch suspensions of different volume fractions: 33\% (\textcolor{red}{$\blacksquare$}), 35\% (\textcolor{blue}{$\blacksquare$}), and 38\%~({\scriptsize$\blacksquare$}). Dashed (continuous) lines are respectively Newtonian (shear-thickening) regimes.}  \label{fig3}
                    \end{figure}

As emphasized, the breaking of axisymmetry is the reason that the deformation of the impinging drop is more appropriately described by the height of the drop $h$ rather than by its spreading diameter. As a result, its maximal deformation is here characterized by the plateau value $\varepsilon_m$ (minimal dimensionless height when the \emph{frozen} state is reached). For both cornstarch and PS spheres suspensions, the influence of the impact speed $U_0$ is studied by systematically varying the drop release height. Snapshots of \emph{frozen} drops (at maximal deformation) for three impact velocities are shown in Fig. \ref{fig1}. Strikingly, the drop maximal deformation turns out to be completely independent of the impact speed over a wide range of $U_0$. Again, this finding is drastically different from the case of Newtonian fluids. Given the relatively large viscosity of the suspensions (at low shear rate, $\eta \sim 50-100$~mPa.s, see Fig.\ref{fig3}), viscous forces will dominate over surface tension and the maximal deformation will be given by a balance between the drop initial kinetic energy (before impact) and viscously-dissipated energy (during the spreading phase) \cite{Clanet2004,Lagubeau2012, Lastakowski2013}. Expressed in terms of $\varepsilon_m$, the resulting prediction for Newtonian liquids reads
 
\begin{equation}
\varepsilon_m \propto Re^{-2/5},
\label{eq:newtonian}
\end{equation}
where $Re=\rho U_0 D_0/\eta$ is the drop impact Reynolds number. 

Clearly, the impact velocity-independence observed for suspensions differs from the variation in $U_0^{-2/5}$ found for Newtonian viscous fluids. Note that velocity-independence is observed for both types of employed particles suspensions, indicating a generic mechanism that is irrespective of chemical interactions and particle shape. 

The influence of the other control parameters $D_0$ and $\phi_0$ is shown in Fig. \ref{fig2}. For a given concentration in cornstarch particles $\phi_0$, the dimensionless minimal height $\varepsilon_m$ is a decreasing function of the initial drop diameter $D_0$. Clearly, the experimental findings are different from the Newtonian result $D_0^{-2/5}$, and are more accurately described by a $D_0^{-2/3}$ scaling that will be derived below. As intuitively expected, more concentrated suspensions exhibit higher minimal thickness $\varepsilon_m$ (smaller deformations). 

These observations reveal an obvious effect of the non-Newtonian properties of concentrated suspensions on the drop impact dynamics. We thus need to incorporate the actual mechanical stress response of these complex fluids and revisit the classical scaling arguments. The shear rheology of cornstarch suspensions is studied using a cylindrical Couette rheometer (Anton Paar Rheolab  QC, CC39 geometry) and the resulting flow curves are shown in Fig. \ref{fig3}. All measurements are taken again after mixing the suspensions with a spatula to provide an homogeneous and reproducible initial state. A time evolution of the measured shear-stress is observed and suggest that particle migration occurs and the suspensions quickly become inhomogeneous in local concentration (classically, particle migrate away from the inner cylinder in cylindrical geometries \cite{Boyer2011,Ovarlez2006}). Therefore, to ensure this dynamical process stays negligible (as it can be expected on the very short timescale of drop impacts), the reported values in Fig. \ref{fig3} are taken at early times of shear (i.e. averaged over the first ten rotations). For all cornstarch concentrations, the flow curves exhibit two regimes: (i) at low shear rates, the shear stress $\tau$ is linear in shear rate $\dot\gamma$ (Newtonian behavior) with an effective viscosity depending solely on the particle volume fraction; (ii) at high shear rates, the shear stress is found to be quadratic in the shear rate. This second regime corresponds to the shear-thickening regime (as the effective viscosity $\eta_{eff}\equiv\tau/\dot\gamma \propto \dot\gamma$ is an increasing function of the shear rate), which is a well-known effect in cornstarch suspensions, and more generally in particulate systems. The shear rheology of PS spheres suspensions has been extensively studied \cite{Fall2010} and has been shown to exhibit the same quadratic regime at high shear rates: as already shown by \cite{Brown2009}, rheology of cornstarch and spherical particles suspensions are thus found to be very similar. An estimate of the minimum shear rate $\dot\gamma_0$ experienced by an impacting drop is given by $\dot\gamma_0= U_0/D_0$ which is always larger than $10^2\, \mathrm{s}^{-1}$ under our experimental conditions. Therefore, our drop impact experiments only probe this shear-thickening regime.

The quadratic dependence observed here suggests a regime where inertia of the particles dominates over viscous forces at the particle scale. This has already been observed for concentrated non-Brownian (i.e. large particles of size $d$) suspensions and is known as \emph{Bagnoldian} scaling, first proposed for granular media from  dimensional arguments. This continuous transition to shear-thickening also indicates that the volume fractions investigated here are sufficiently below the jamming point to avoid more severe shear-thickening effects, for which linking any global rheology to local constitutive laws remained controversial \cite{Fall2010,Brown2009}. We therefore fit the quadratic regime to the form
\begin{equation}
\tau = \kappa(\phi) \dot\gamma^2,
\label{eq:rheology}
\end{equation}
and revisit the scaling argument for drop impact, taking into account this shear thickening rheology. They idea is that the initial kinetic energy of the drop is annihilated by viscous dissipation in the drop. The rate of dissipation per unit volume is computed as $\tau \dot\gamma =  \kappa(\phi) \dot\gamma^3$, where the strain rate $\dot\gamma$ can be estimated as $\propto U_0/(\epsilon_m D_0)$ during impact. Since the impact time scales as $\propto D_0/U_0$, the total amount of dissipated energy becomes

\begin{equation}
\frac{D_0}{U_0} \int d^3 \mathbf{r} \, \tau \dot\gamma \, \propto \, 
\kappa(\phi) \left(\frac{U_0}{\varepsilon_m D_0}\right)^3 \left(\frac{D_0}{U_0}\right) D_0^3.
\end{equation}
Equating this to the kinetic energy, $\rho D_0^3 U_0^2$, we obtain the maximum deformation

\begin{eqnarray}
\varepsilon_m &\propto& \left(\frac{\kappa(\phi)}{\rho}\right)^{1/3} D_0^{-2/3}.
\label{eq:scaling}
\end{eqnarray}
In agreement with experimental observations, the maximal deformation is found to be independent of the impact velocity $U_0$ in stark contrast to Newtonian fluids. Defining a non-Newtonian Reynolds number would have also lead to the same statement: for complex fluids which follow $\tau= K \dot\gamma^n$, the generalized Reynolds number reads $Re_n=\rho U_0^{2-n} D_0^n/K$ \cite{Luu2009}. In the present case, the shear-thickening exponent is $n=2$ which cancels any dependence of the generalized Reynolds number with the velocity $U_0$ and then predicts the same observed behavior. 
                    
                        \begin{figure}[htp]
                    \center{\includegraphics[width=8.6cm]{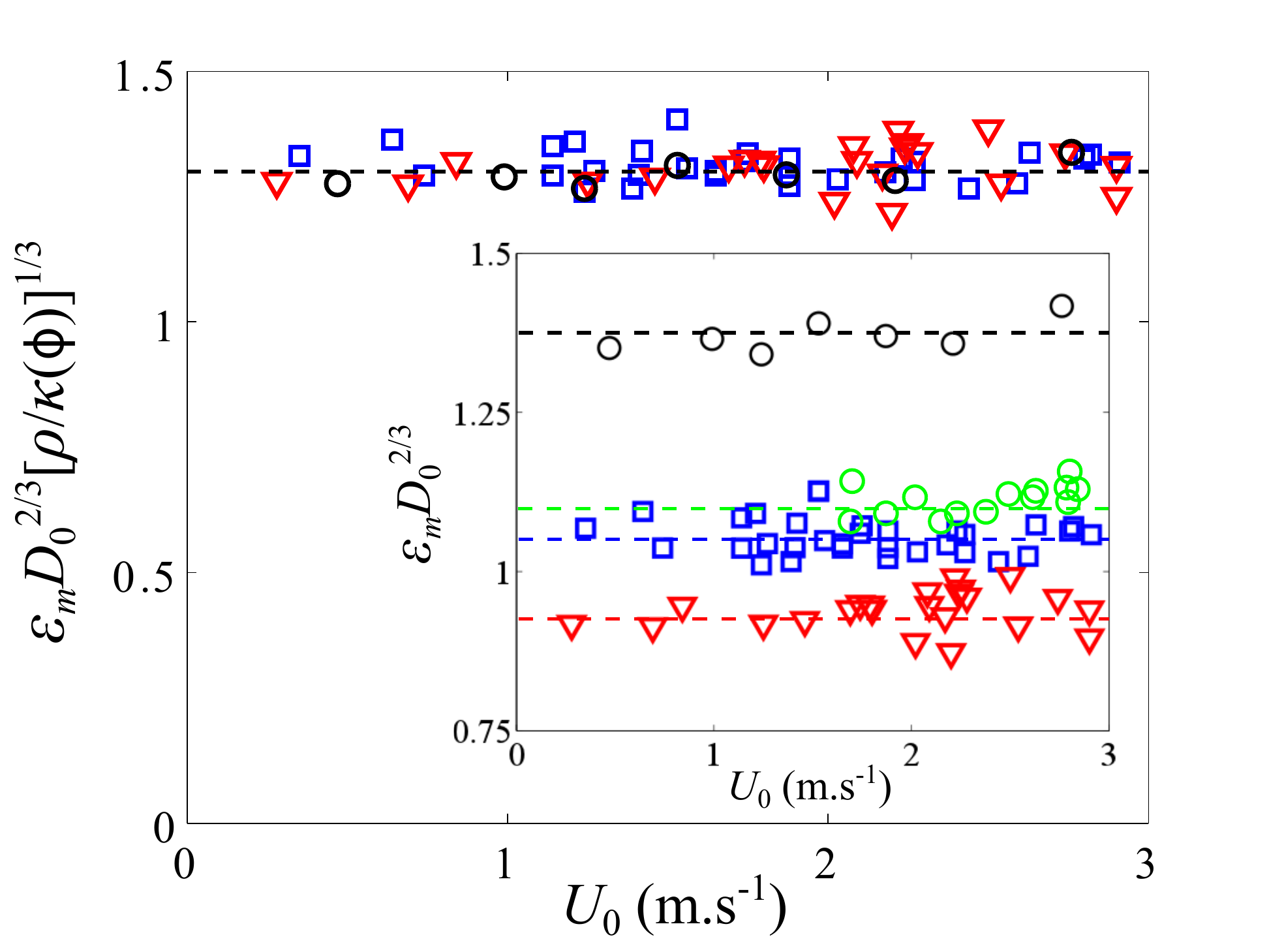}}
                    \caption{(color online) Rescaled minimal height $\varepsilon_m D_0^{2/3} [\rho/\kappa(\phi)]^{1/3}$ as a function of impact velocity $U_0$ for different cornstarch volume fractions: 33\% (\textcolor{red}{$\bigtriangledown$}), 35\% (\textcolor{blue}{$\square$}), and 38\% ({\scriptsize$\bigcirc$}). 
                    Inset: Rescaled minimal height $\varepsilon_m D_0^{2/3}$.}  \label{fig4}
                    \end{figure}
                    
The dependence of the drop diameter on the minimal thickness $\varepsilon_m$ obtained in Eq. (\ref{eq:scaling}) is indeed found to be in very good agreement with our observations (Fig. \ref{fig2}). This allows us to recast the experimental results to further emphasize the independence of the spreading on the impact speed. In the inset of Fig. \ref{fig4}, we find that $\varepsilon_m D_0^{2/3}$ is indeed perfectly constant over the range of investigated impact speeds (one order of magnitude) for all suspensions. We further test our scaling argument by using  the rheology $\kappa(\phi)$ for three different volume fractions of cornstarch, as obtained from independent rheological measurement. Indeed, all experimental data can be scaled onto a single horizontal line (Fig. \ref{fig4}), suggesting a universal numerical prefactor in Eq. (\ref{eq:scaling}). As a result, the present drop impact experiments of cornstarch suspensions are shown to be accurately modeled by
\begin{equation}
\varepsilon_m = 1.3 \ Re_{ST}^{-1/3},
\label{eq:shearthick}
\end{equation}
where $Re_{ST} = \rho D_0^2/\kappa(\phi)$ is a generalized Reynolds number for quadratic shear-thickening fluids ($n=2$).

As a conclusion, the drop impact of concentrated non-Brownian suspensions onto a solid surface exhibits clear differences with Newtonian fluids: both the dynamics and final state show unusual behaviors, the most striking finding being the velocity-independence of the maximal deformation. Such observation is of primary interest for numerous applications where full control of the drop size after impact is required: in this context, suspensions are an interesting class of materials for which maximal deformation depends only on initial drop size.

Revisiting classical energetic arguments while integrating the complex rheology of shear-thickening suspensions leads to successful scaling laws. In particular, the quadratic constitutive law (consistent with well-known \emph{Bagnoldian} inertial regime) is shown to be a key point to rationalize the observed velocity-independent maximal deformation. As already shown for shear rheology, cornstarch suspensions show the same behavior as monodisperse hard-spheres systems, suggesting chemical interactions between particles are rather insignificant in controlling macroscopic response for these systems. Finally, quantitatively modeling of all our experimental data with one single equation makes it possible to directly infer the rheological quantity $\kappa(\phi)$ from drop impact observations, thus opening the way to develop a \emph{drop impact rheometry}.

We thank Enrique Sandoval Nava and Stef Carelsen for help with rheological measurements. This work was funded by the Dutch Polymer Institute under the 'Inkjet-Priniting of Suspensions' project.


\end{document}